\documentclass[preprint,12pt,preprintnumbers,amsmath,amssymb,floatfix,endfloats*]{revtex4}

\usepackage{graphicx}
\usepackage{dcolumn}
\usepackage{bm}
\usepackage{amsfonts}

\DeclareMathAlphabet{\mathsfsl}{OT1}{cmr}{bx}{it}
\begin{document}
\title{Slip boundary conditions for shear flow of polymer melts past atomically flat surfaces}
\author{Anoosheh Niavarani and Nikolai~V.~Priezjev}


\affiliation{Department of Mechanical Engineering, Michigan State
University, East Lansing, Michigan 48824}
\date{\today}
%
\begin{abstract}

Molecular dynamics simulations are carried out to investigate the
dynamic behavior of the slip length in thin polymer films confined
between atomically smooth thermal surfaces. For weak wall-fluid
interactions, the shear rate dependence of the slip length acquires
a distinct local minimum followed by a rapid growth at higher shear
rates. With increasing fluid density, the position of the local
minimum is shifted to lower shear rates. We found that the ratio of
the shear viscosity to the slip length, which defines the friction
coefficient at the liquid/solid interface, undergoes a transition
from a nearly constant value to the power law decay as a function of
the slip velocity. In a wide range of shear rates and fluid
densities, the friction coefficient is determined by the product of
the value of surface induced peak in the structure factor and the
contact density of the first fluid layer near the solid wall.

\end{abstract}

\pacs{68.08.-p, 83.80.Sg, 83.50.Rp, 47.61.-k, 83.10.Rs}


\maketitle

\section{Introduction}

The fluid flow in microfluidic channels can be significantly
influenced by liquid slip at the solid boundary because of the large
surface to volume ratio~\cite{MicroNano}. The Navier model for the
partial slip boundary conditions relates the fluid slip velocity to
the tangential viscous stress at the wall via the friction
coefficient, which is defined as the ratio of the shear viscosity to
the slip length. Geometrically, the slip length is determined as a
distance from the solid wall where the linearly extrapolated fluid
velocity profile vanishes. Experimental studies have demonstrated
that for atomically smooth surfaces the slip length is relatively
large for non-wetting liquid/solid
interfaces~\cite{Churaev84,LegerPRL93,Charlaix01}, polymeric
fluids~\cite{Archer98,MackayVino} and high shear
rates~\cite{Mackay00,Granick01,Breuer03}. By contrast, surface
roughness even on the submicron length scale strongly reduces the
degree of slip for both polymeric~\cite{Archer03} and
Newtonian~\cite{Granick02,Vinograd06,Leger06} fluids. However,
experimentally it is often difficult to isolate and, consequently,
to analyze the effects of wettability, surface roughness and shear
rate on slip boundary conditions because of the small scales
involved.

During the past two decades, a number of molecular dynamics (MD)
simulations~\cite{KB89,Thompson90,Barrat99fd,Barrat99,Tanner99,Cieplak01,Priezjev04,Attard04,Priezjev05,Priezjev07}
have been performed to investigate the dependence of the slip length
on the structure of dense fluids near atomically flat surfaces. For
simple fluids, it has been well established that the slip length
correlates inversely with the wall-fluid interaction energy and the
surface induced order in the first fluid
layer~\cite{Thompson90,Barrat99fd,Priezjev07}. The strong wall-fluid
coupling and commensurate structures of the liquid/solid interface
lead to the no-slip boundary condition for monoatomic
fluids~\cite{Thompson90,Attard04}. The interfacial slip is enhanced
in the flow of polymer melts due to the higher shear viscosity and
reduced molecular ordering near the flat
wall~\cite{Thompson95,dePablo96,Koike98,Priezjev04}. The magnitude
of the slip length in the flow of polycarbonate blends past
atomically smooth nickel surfaces is determined by the competition
between the apparent slip of the adsorbed polymer layer and its
thickness~\cite{Andrienko05}. The slip boundary conditions for the
polymer melt flow over the brush can be controlled by changing the
density of the end-grafted polymer layer~\cite{Binder06}. Finally,
the degree of slip at the interface between immiscible polymers is
larger for longer chain molecules and greater repulsion between
unlike species~\cite{Barsky01}.

Until ten years ago, the slip length in the Navier model was assumed
to be independent of shear rate. In the seminal MD study by Thompson
and Troian~\cite{Nature97} on shear flow of simple fluids past
atomically flat rigid walls, it was found that the slip length
increases nonlinearly with the shear rate. Upon increasing the
surface energy, a gradual transition from the power law
behavior~\cite{Nature97} to the linear rate dependence was recently
reported for weak wall-fluid interactions and incommensurate
structures of the liquid/solid interface~\cite{Priezjev07}. In a
wide range of shear rates and surface energies, the slip length in
simple fluids is determined by a function of the in-plane structure
factor, contact density and temperature of the first fluid layer
near the wall~\cite{Priezjev07}. The rate-dependent slip length at
the interface between a polymer melt composed of short linear chains
and a crystalline substrate is also well described by the power law
function used to fit the data for simple
fluids~\cite{Priezjev04,Nature97}. Whether these boundary conditions
hold for higher molecular weight polymers or rough surfaces is an
open question that has to be addressed for the correct
interpretation of the experimental results and modeling fluid flows
in microfluidic channels.

In this paper, the effects of shear rate and fluid density on slip
boundary conditions for the flow of a polymer melt past a
crystalline wall are investigated by molecular dynamics simulations.
We will show that the rate-dependent slip length reaches a local
minimum and then increases rapidly at higher shear rates. In the
shear thinning regime, the friction coefficient at the liquid/solid
interface follows a power law decay as a function of the slip
velocity. In a wide range of shear rates and polymer densities, the
ratio of the slip length to the shear viscosity is determined by the
surface induced structure in the first fluid layer. These results
suggest that the correlation between the fluid structure near the
flat wall and the slip length is determined by a universal
combination of parameters for both polymer melts and simple
fluids~\cite{Priezjev07}.

This paper is organized as follows. The molecular dynamics
simulation model and parameter values are described in the next
section. The results for the rate dependence of the slip length and
the analysis of the fluid structure near the solid wall are
presented in Section~\ref{sec:Results}. The conclusions are given in
the last section.

\section{Details of molecular dynamics simulations}
\label{sec:Model}

\begin{figure}[t]
\vspace*{-3mm}
\includegraphics[width=7.0cm,height=6.0cm,angle=0]{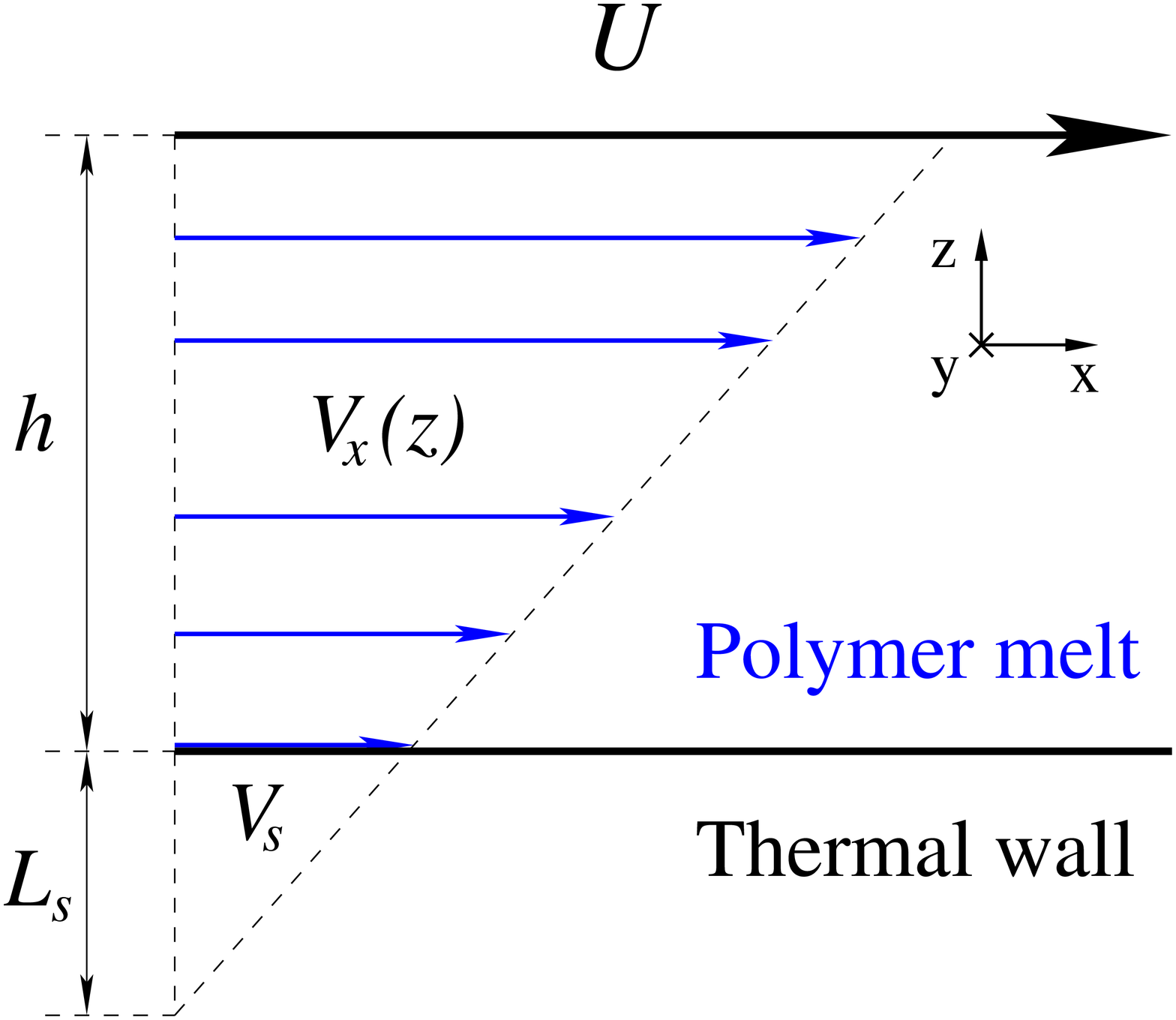}
\caption{(Color online) A schematic representation of the
steady-state shear flow in the Couette cell. The upper wall is
translated with a constant velocity $U$ in the $\hat{x}$ direction.
The fluid slip velocity is determined from the relation
$V_s\,{=}\,\,\dot{\gamma}L_s$, where $\dot{\gamma}$ is the slope of
the velocity profile and $L_s$ is the slip length.}
\label{schematic}
\end{figure}

The polymer melt is confined between two atomically flat walls and
is subject to planar shear in the $\hat{x}$ direction (see
Fig.\,\ref{schematic} for a schematic representation). The fluid
monomers interact through the pairwise Lennard-Jones (LJ) potential
\begin{equation}
V_{LJ}(r)\!=4\,\varepsilon\,\Big[\Big(\frac{\sigma}{r}\Big)^{12}\!-\Big(\frac{\sigma}{r}\Big)^{6}\,\Big],
\end{equation}
with a cutoff distance $r_c\!=\!2.5\,\sigma$, where $\varepsilon$
and $\sigma$ represent the energy and length scales of the fluid
phase. The total number of monomers is $N_{f}\!=6000$. The
wall-fluid interaction parameters of the LJ potential are set to
$\varepsilon_{\rm wf}\,{=}\,0.9\,\varepsilon$ and $\sigma_{\rm
wf}\,{=}\,\sigma$ throughout the study. The polymer melt is modeled
as a collection of bead-spring chains of $N\,{=}\,20$ LJ monomers
interacting through the FENE (finitely extensible nonlinear elastic)
potential~\cite{Bird87}
\begin{equation}
V_{FENE}(r)=-\frac{k}{2}\,r_{\!o}^2\ln[1-r^2/r_{\!o}^2],
\end{equation}
with Kremer and Grest parameters
$k\,{=}\,30\,\varepsilon\sigma^{-2}$ and
$r_{\!o}\,{=}\,1.5\,\sigma$~\cite{Kremer90}. This choice of
parameters prevents polymer chains from unphysical crossing or bond
breaking~\cite{Grest86}.

The system was coupled to an external heat bath through a Langevin
thermostat~\cite{Grest86} with a random force and a damping term
with a small value of the friction coefficient
$\Gamma\,{=}\,1.0\,\tau^{-1}$~\cite{GrestJCP04}. The thermostat was
only applied to the equation of motion for a fluid monomer along the
$\hat{y}$ axis to avoid a bias in the shear flow
direction~\cite{Thompson90}. The equations of motion for the
$\hat{x}$, $\hat{y}$ and $\hat{z}$ components are given by
\begin{eqnarray}
\label{Langevin_x}
m\ddot{x}_i & = & -\sum_{i \neq j} \frac{\partial V_{ij}}{\partial x_i}\,, \\
\label{Langevin_y}
m\ddot{y}_i + m\Gamma\dot{y}_i & = & -\sum_{i \neq j} \frac{\partial V_{ij}}{\partial y_i} + f_i\,, \\
\label{Langevin_z}
m\ddot{z}_i & = & -\sum_{i \neq j} \frac{\partial V_{ij}}{\partial z_i}\,, %
\end{eqnarray}
where $f_i$ is a random force with zero mean and variance $\langle
f_i(0)f_j(t)\rangle\,{=}\,\,2mk_BT\Gamma\delta(t)\delta_{ij}$, the
temperature is $T\,{=}\,1.1\,\varepsilon/k_B$ and $k_B$ is the
Boltzmann constant. The equations of motion for fluid monomers and
wall atoms were integrated using the fifth-order gear-predictor
algorithm~\cite{Allen87} with a time step $\triangle
t\,{=}\,0.002\,\tau$, where $\tau\!=\!\sqrt{m\sigma^2/\varepsilon}$
is the characteristic time of the LJ potential.

\begin{table}[b]
\caption{The fluid pressure $P$ at equilibrium, i.e., with the
stationary upper wall, and the channel height $h$ (defined as a
distance between the inner fcc planes) as a function of the fluid
density.}
 \vspace*{3mm}
 \begin{ruledtabular}
 \begin{tabular}{r r r r r r r}
     $\rho\,[\,\sigma^{-3}]$ & $0.86$ & $0.88$ & $0.91$ & $0.94$ & $1.00$ & $1.02$
     \\ [3pt] \hline 
     $P\,[\,\varepsilon\,\sigma^{-3}]$ &    $0.0$ &  $0.5$ & $1.0$ & $2.0$ & $4.0$ & $5.0$
     \\ [3pt]
     $h\,[\sigma]$   &  $28.93$ & $28.18$  & $27.27$ & $26.32$ & $24.85$ & $24.36$
     \\ [2pt]
 \end{tabular}
 \end{ruledtabular}
 \label{tabela}
\end{table}

Each wall was constructed out of $576$ LJ atoms, which formed two
(111) layers of the face-centered cubic (fcc) lattice with the
density $\rho_w\,{=}\,1.40\,\sigma^{-3}$. The corresponding
nearest-neighbor distance between equilibrium positions of the wall
atoms in the $xy$ plane is $d\,{=}\,1.0\,\sigma$. The dimensions of
the Couette cell ($xyz$) were held fixed to $20.86\,\sigma \times
12.04\,\sigma\times h$, where $h$ is the channel height (see
Table\,\ref{tabela}). The MD simulations were performed at a
constant density ensemble. The fluid density is defined as a ratio
of the total number of monomers to the volume accessible to the
fluid, i.e., $0.5\,\sigma_{\rm w}$ away from the fcc lattice planes
in the $\hat{z}$ direction. The range of fluid densities considered
in this study is $0.86\leqslant\rho\,\sigma^3\leqslant1.02$. The
pressures of the system at equilibrium are given in
Table\,\ref{tabela} for each value of the fluid density. Periodic
boundary conditions were employed along the $\hat{x}$ and $\hat{y}$
directions parallel to the confining walls.

The wall atoms were attached to their equilibrium fcc lattice
positions by the harmonic spring with the potential
$V_{sp}\,{=}\,\frac{1}{2}\,\kappa\,r^2$. The spring stiffness
$\kappa\,{=}\,1200\,\varepsilon/\sigma^2$ is large enough so that
the mean-square displacement of the wall atoms is less than the
Lindemann criterion for melting, i.e., $\langle\delta
u^2\rangle/d^{\,2}\lesssim 0.023$ (see Ref.~\cite{Barrat03}). In the
previous paper on shear flow of simple fluids~\cite{PriezjevJCP}, it
was shown that the slope of the rate-dependent slip length remains
approximately the same for the thermal walls with
$\kappa\,{=}\,1200\,\varepsilon/\sigma^2$ and rigid fcc walls. The
friction term and the random force were also added to the $\hat{x}$,
$\hat{y}$ and $\hat{z}$ components of the wall atoms equations of
motion
\begin{eqnarray}
\label{Langevin_wall_x} m_w\,\ddot{x}_i + m_w\,\Gamma\dot{x}_i & = &
-\sum_{i \neq j} \frac{\partial V_{ij}}{\partial x_i} -
\frac{\partial V_{sp}}{\partial x_i} + f_i\,, \\
\label{Langevin_wall_y} m_w\,\ddot{y}_i + m_w\,\Gamma\dot{y}_i & = &
-\sum_{i \neq j} \frac{\partial V_{ij}}{\partial y_i}
- \frac{\partial V_{sp}}{\partial y_i} + f_i\,, \\
\label{Langevin_wall_z} m_w\,\ddot{z}_i + m_w\,\Gamma\dot{z}_i & = &
-\sum_{i \neq j} \frac{\partial V_{ij}}{\partial z_i} -
\frac{\partial V_{sp}}{\partial z_i} + f_i\,,%
\end{eqnarray}
where the mass of the wall atoms is $m_w\,{=}\,10\,m$ and the
summation is performed only over the fluid monomers within the
cutoff distance $r_c\!\,\,{=}\,\,2.5\,\sigma$.

The simulations began with an equilibration period of about
$5\times10^4\tau$ with both walls being at rest. Then, the
steady-state shear flow was generated by translating the upper wall
with a constant velocity $U$ in the $\hat{x}$ direction, while the
lower wall always remained stationary (see Fig.\,\ref{schematic}).
The fluid velocity profiles were computed by averaging the
instantaneous monomer velocities in the $\hat{x}$ direction within
bins of thickness $\Delta z\!=\!0.2\,\sigma$ for a time interval up
to $4\times10^5\tau$ at the lowest shear rates. The density profiles
were averaged for a time period $5\times10^4\tau$ within bins of
thickness of only $\Delta z\,{=}\,0.01\,\sigma$ to accurately
resolve the fluid structure near the walls~\cite{Priezjev07}. The
maximum value of the Reynolds number is $Re\!\approx\!10.5$, which
was estimated from the maximum fluid velocity difference across the
channel, the shear viscosity, and the channel height. The small
values of the Reynolds number correspond to laminar flow conditions
even at the highest shear rates considered in this study.

\begin{figure}[t]
\includegraphics[width=10.cm,height=7.35cm,angle=0]{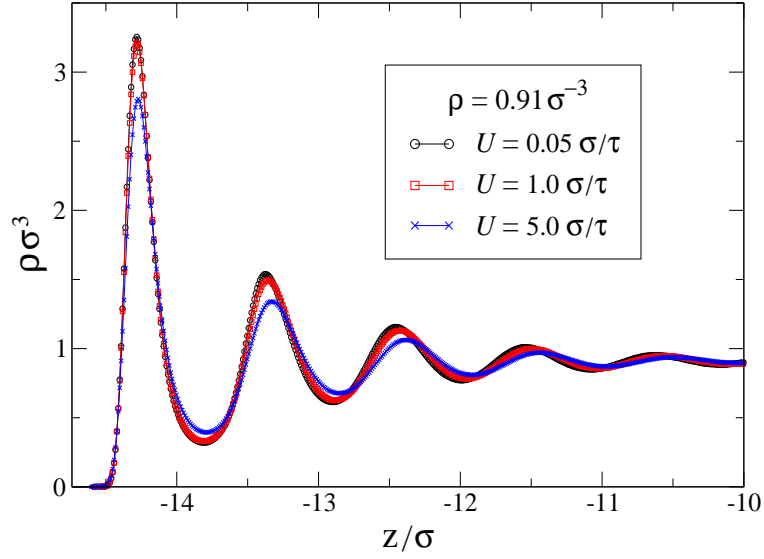}
\caption{(Color online) Averaged fluid density profiles near the
stationary thermal wall with
$\kappa\,{=}\,1200\,\varepsilon/\sigma^2$ and $\varepsilon_{\rm
wf}/\varepsilon\,{=}\,0.9$. The velocities of the upper wall are
tabulated in the inset. The fluid density in the center of the cell
is $\rho\,{=}\,0.91\,\sigma^{-3}$. The left vertical axis denotes
the position of the liquid/solid interface at
$z\,{=}\,-14.74\,\sigma$.} \label{mol_dens}
\end{figure}

\section{Results}
\label{sec:Results}

\subsection{Fluid density and velocity profiles}

The fluid structure in the direction perpendicular to a flat surface
usually consists of several distinct layers on the molecular
scale~\cite{Israel92}. Examples of the averaged monomer density
profiles near the lower thermal wall are presented in
Fig.\,\ref{mol_dens} for three values of the upper wall velocity.
The highest peak is associated with a layer of fluid monomers (each
connected to a polymer chain) in contact with the wall atoms. The
contact density $\rho_c$ corresponds to the maximum value of the
peak. The density oscillations gradually decay away from the wall to
a uniform profile with bulk density $\rho\,{=}\,0.91\,\sigma^{-3}$.
The amplitude of the density oscillations is reduced at higher
values of the slip velocity.



\begin{figure}[t]
\includegraphics[width=10.cm,height=7.35cm,angle=0]{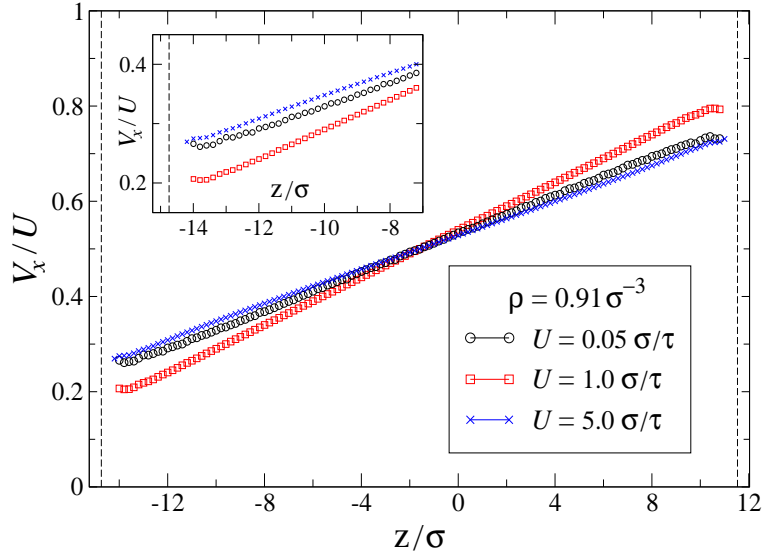}
\caption{(Color online) Average normalized velocity profiles,
$V_{x}/\,U$, for the indicated values of the upper wall velocity $U$
and the fluid density $\rho\,{=}\,0.91\,\sigma^{-3}$. Vertical axes
denote the position of the fcc lattice planes at
$z\,{=}\,-15.24\,\sigma$ and $z\,{=}\,12.03\,\sigma$. The dashed
lines indicate the location of liquid/solid interfaces at a distance
$0.5\,\sigma_{\rm w}$ away from the fcc lattice planes, i.e.,
$z\,{=}\,-14.74\,\sigma$ and $z\,{=}\,11.53\,\sigma$. The inset
shows the same data near the lower wall.} \label{shear_velo}
\end{figure}

Figure\,\ref{shear_velo} shows the averaged velocity profiles for
the same values of the upper wall speed as in Fig.\,\ref{mol_dens}
and the fluid density $\rho\,{=}\,0.91\,\sigma^{-3}$. For a given
$U$, the relative slip velocity is the same at the upper and lower
walls. Surprisingly, the scaled slip velocity is smaller at the
intermediate speed of the upper wall $U\,{=}\,\,1.0\,\sigma/\tau$.
Because of the thermal fluctuations, longer averaging times (up to
$4\times10^5\tau$) were required to accurately resolve the fluid
velocity profile for $U\,{=}\,\,0.05\,\sigma/\tau$. A small
curvature in the velocity profile is evident for the largest
velocity of the upper wall $U\,{=}\,\,5.0\,\sigma/\tau$. The
deviation from the linearity might be related to the heating up of
the fluid near the walls (see discussion below).
The slope of the velocity profile (used for calculation of the slip
length and the shear rate) was computed from the central part of the
velocity profile excluding the data from the region of about
$2\,\sigma$ near the walls.

We comment that at small shear rates,
$\dot{\gamma}\tau\lesssim0.01$, and higher fluid densities,
$\rho\gtrsim1.04\,\sigma^{-3}$, the fluid velocity profiles acquired
a pronounced curvature within a distance of about $4\,\sigma$ from
the walls. The linearity of the fluid velocity profiles is restored
at higher shear rates. This effect will be examined in a separate
study.

\subsection{Rate dependence of the slip length and shear viscosity}

In the MD study by Thompson and Troian~\cite{Nature97} on shear flow
of Newtonian liquids past atomically flat surfaces, the slip length
was found to increase nonlinearly with the shear rate. The data for
different wall densities and weak wall-fluid interactions were
fitted well by the power law function
\begin{equation}
L_s(\dot{\gamma})=L_s^{o}~(1-{\dot{\gamma}}/{\dot{\gamma}}_c)^{-0.5},
\label{nature}
\end{equation}
where $\dot{\gamma_c}$ and $L_s^{o}$ are the parameters used for
normalization. The rate-dependent slip length in the Poiseuille flow
of simple fluids was also well described by the function
[Eq.\,(\ref{nature})] for incommensurate structures of the
liquid/solid interface and weak surface energy~\cite{Priezjev07}.

\begin{figure}[t]
\includegraphics[width=10.cm,height=7.35cm,angle=0]{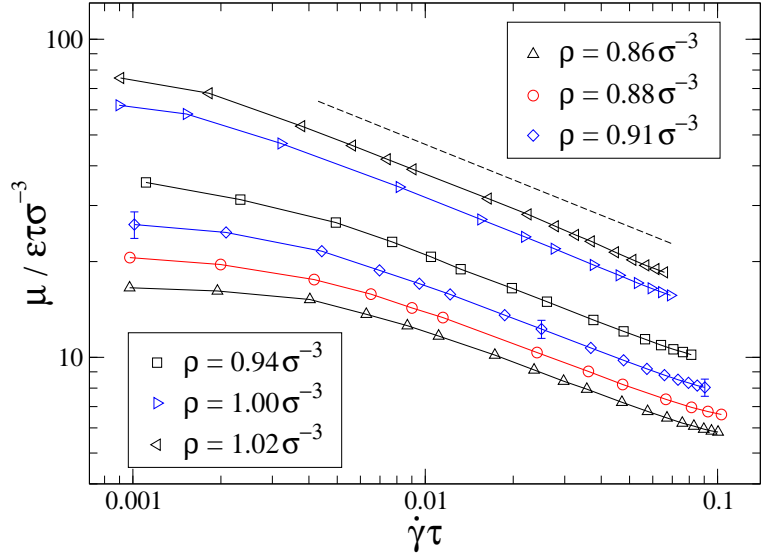}
\caption{(Color online) Behavior of the fluid viscosity
$\mu/\,\varepsilon\tau\sigma^{-3}$ as a function of shear rate for
the tabulated values of the polymer density. The straight dashed
line with a slope $-0.37$ is shown for reference. Solid curves are a
guide for the eye.} \label{visc_shear_all}
\end{figure}

The nonlinear behavior of the slip length can be interpreted in
terms of the velocity-dependent shear stress at the
wall~\cite{Robbins97}. By definition, the fluid slip velocity $V_s$
is related to the slip length as $V_s\,{=}\,\,\dot{\gamma}L_s$. On
the other hand, the shear stress through any plane parallel to the
walls is the same in the steady-state shear flow, and it is equal to
the fluid viscosity multiplied by the shear rate, i.e.,
$\sigma_{xz}\,{=}\,\,\mu\dot{\gamma}$. At the liquid/solid interface
the viscous shear stress becomes $\sigma_{xz}\,{=}\,\,kV_s$, where
$k$ is the friction coefficient per unit area. The functional form
given by Eq.\,(\ref{nature}) implies that the friction coefficient,
$k\,{=}\,\,\mu/L_s$, can be expressed as a function of the slip
velocity as follows:
\begin{equation}
k(V_s)=C_1( \sqrt{C_2+V^{2}_s}-V_s), \label{nature_friction}
\end{equation}
where $C_1=\mu/2\dot{\gamma}_c(L^{o}_s)^2$ and
$C_2=(2L^{o}_s\dot{\gamma}_c)^2$, and $\mu$ is the shear-independent
viscosity~\cite{Nature97,Priezjev04}.

In the previous MD study~\cite{Priezjev04}, it was shown that the
slip length increases monotonically with shear rate for polymer
melts with short linear chains $N\leqslant16$. The data for the slip
length were well fitted by the power law function given by
Eq.\,(\ref{nature}) for shear rates exceeding
$\dot{\gamma}\gtrsim5\times10^{-3}\,\tau^{-1}$. Although the fluid
viscosity exhibited a slight shear thinning behavior for chain
lengths $N\geqslant8$, no minimum in the rate dependence of the slip
length was observed. In the present study, the shear thinning effect
is more pronounced because the polymer melt consists of longer
chains and the simulations are performed at higher fluid densities.

\begin{figure}[t]
\includegraphics[width=10.cm,height=7.35cm,angle=0]{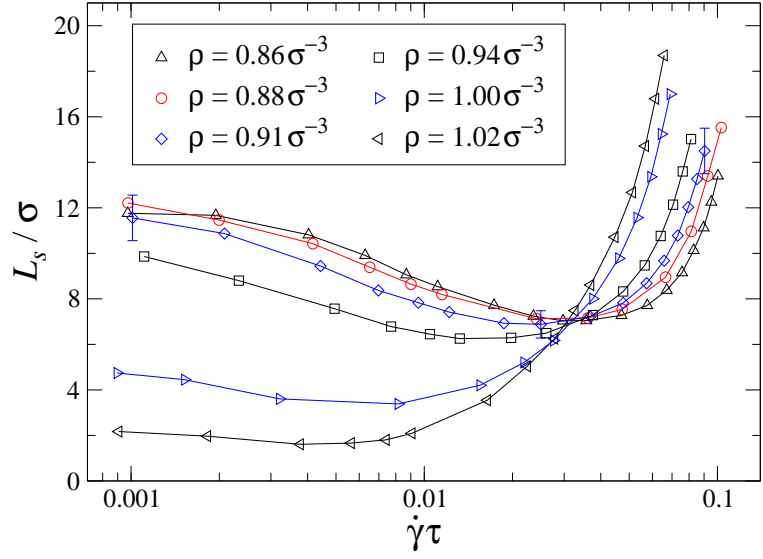}
\caption{(Color online) Slip length $L_s/\sigma$ as a function of
the shear rate for the polymer melt with linear chains $N\,{=}\,20$.
Solid curves are a guide for the eye.}\label{shear_ls_all}
\end{figure}

Figure\,\ref{visc_shear_all} shows the rate behavior of the fluid
viscosity, which was computed from the Kirkwood relation for the
shear stress~\cite{Bird87} in steady-state shear flow. In the range
of accessible shear rates, the transition from the Newtonian regime
to the shear thinning is more evident at lower fluid densities. At
higher shear rates, the bulk viscosity approximately follows a power
law decay with the exponent $-0.37$ (see
Fig.\,\ref{visc_shear_all}). A similar shear thinning behavior of
polymer melts was reported in the previous MD
studies~\cite{dePablo95,Todd04}. The error bars are larger at lower
shear rates because of an increasing role of thermal
fluctuations~\cite{XudePablo97}.

The dynamic response of the slip length with increasing shear rate
is shown in Fig.\,\ref{shear_ls_all} for the indicated values of the
polymer density. The rate dependence of the slip length exhibits a
distinct local minimum, which is followed by a rapid increase at
higher shear rates. The minimum is shifted to lower shear rates and
becomes more shallow at higher fluid densities. The slip length
tends to saturate to a constant value at the lowest shear rates for
the fluid density $\rho\,{=}\,0.86\,\sigma^{-3}$. Typical error bars
are included for three points corresponding to the fluid velocity
profiles shown in Fig.\,\ref{shear_velo}. In the range of shear
rates from the local minimum up to the highest $\dot{\gamma}$, the
data for the slip length (as well as the ratio $L_s/\mu$) cannot be
fitted by the power law function given by Eq.\,(\ref{nature}). For
each value of the fluid density, the highest shear rate corresponds
to the maximum sustainable shear stress at the
interface~\cite{Priezjev07}.

\begin{figure}[t]
\includegraphics[width=10.cm,height=7.35cm,angle=0]{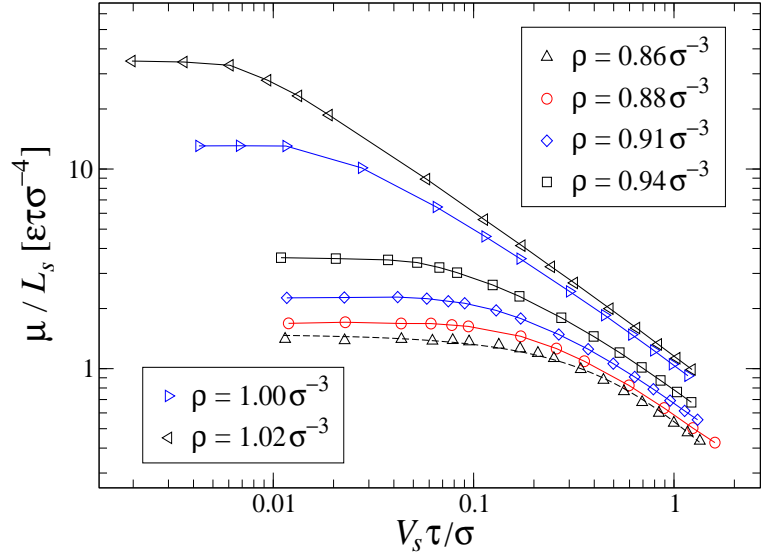}
\caption{(Color online) Log-log plot of the friction coefficient,
$k\,{=}\,\,\mu/L_s$, as a function of the slip velocity for the
indicated values of the fluid density. The dashed curve is the best
fit to Eq.\,(\ref{nature_friction}) with $C_1=1.77\,m\sigma^{-3}$
and $C_2=0.71\,\sigma^2\tau^{-2}$. Solid curves are a guide for the
eye.} \label{friction_shear_all}
\end{figure}

The data for the rate-dependent slip length and fluid viscosity are
replotted in Fig.\,\ref{friction_shear_all} in terms of the friction
coefficient at the liquid/solid interface and the slip velocity. The
constant value of the friction coefficient at small slip velocities
indicates that the ratio of the slip length and viscosity is
rate-independent. At higher fluid densities (or pressures), the
friction coefficient is larger, and the transition to the power law
regime is shifted to smaller slip velocities. The transition point
approximately corresponds to the location of the minimum in the
shear rate dependence of the slip length shown in
Fig.\,\ref{shear_ls_all}. The data for the lowest fluid density
$\rho\,{=}\,0.86\,\sigma^{-3}$ can be fairly well fitted by
Eq.\,(\ref{nature_friction}) [\,see
Fig.\,\ref{friction_shear_all}\,], but the fit becomes worse at
higher fluid densities (not shown). The simulation results indicate
that the nonmonotonic rate behavior of the slip length for polymer
melts can be ascribed to the velocity dependence of the friction
coefficient, which undergoes a transition from a nearly constant
value to a decreasing function of the slip velocity.

\begin{figure}[t]
\includegraphics[width=10.cm,height=7.35cm,angle=0]{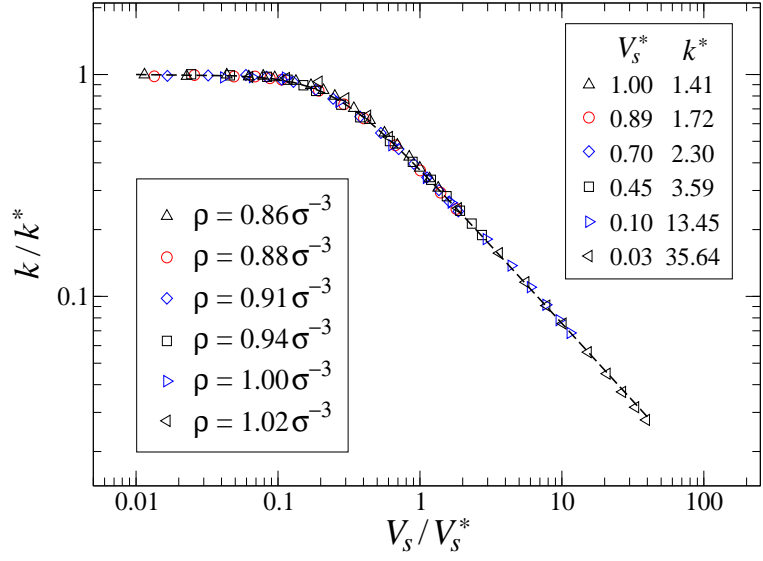}
\caption{(Color online) Master curve for the friction coefficient as
a function of the slip velocity. The same data as in
Fig.\,\ref{friction_shear_all}. The values of the friction
coefficient $k^{\ast}\,[\,\varepsilon\tau\sigma^{-4}]$ and the slip
velocity $V^{\ast}_s\,[\,\sigma/\tau\,]$ used for normalization of
the data are shown in the inset. The dashed line
$y=[1+(4x)^2]^{-0.35}$ is plotted for reference.}
\label{friction_shear_all_master}
\end{figure}

Furthermore, the data for different fluid densities can be collapsed
onto a single master curve by scaling both axes (see
Fig.\,\ref{friction_shear_all_master}). The values of the
normalization parameters for the friction coefficient and the slip
velocity are listed in the inset. The decay at large slip velocities
approximately follows the power law function with the exponent
$-0.7$ (see Fig.\,\ref{friction_shear_all_master}). The onset of the
power law decay depends on the polymer density and might occur when
the time scale of the in-plane diffusion of the fluid monomers over
the lattice spacing $d$ is comparable to $d/V_s$. We also note that
the data for the viscosity and slip length at the interface between
a polymer melt (with linear chains $N\,{=}\,20$ and density
$\rho\,{=}\,0.88\,\sigma^{-3}$) and a rigid fcc wall (with density
$\rho_w\,{=}\,1.94\,\sigma^{-3}$ and $\varepsilon_{\rm
wf}\,{=}\,\varepsilon$) can be normalized to follow the master curve
shown in Fig.\,\ref{friction_shear_all_master}~\cite{Priezjev08}.

\subsection{Analysis of the fluid structure in the first layer}

The surface induced order in the adjacent fluid layer is described
by the structure factor at the reciprocal lattice vectors of the
crystal wall~\cite{Thompson90}. The structure factor is defined as
\begin{equation}
S(\mathbf{k})=\frac{1}{N_{\ell}}\,\,\Big|\sum_j
e^{i\,\mathbf{k}\cdot\mathbf{r}_j}\Big|^2,
\end{equation}
where $\mathbf{r}_j\,{=}\,(x_j,y_j)$ is the position vector of the
$j$-th monomer and the summation is performed over $N_{\ell}$
monomers within the first layer. The friction force between a layer
of adsorbed monomers and a solid substrate is proportional to the
peak value of the in-plane structure factor estimated at the first
reciprocal lattice vector~\cite{Smith96,Tomassone97}. Previous MD
studies of simple fluids have also reported a strong correlation
between the slip length (which is inversely proportional to the
friction coefficient at the liquid/solid interface) and the surface
induced order in the first fluid
layer~\cite{Thompson90,Barrat99fd,Priezjev07}.

\begin{figure}[t]
\vspace*{-5mm}%
\includegraphics[width=10.0cm,height=8.5cm,angle=0]{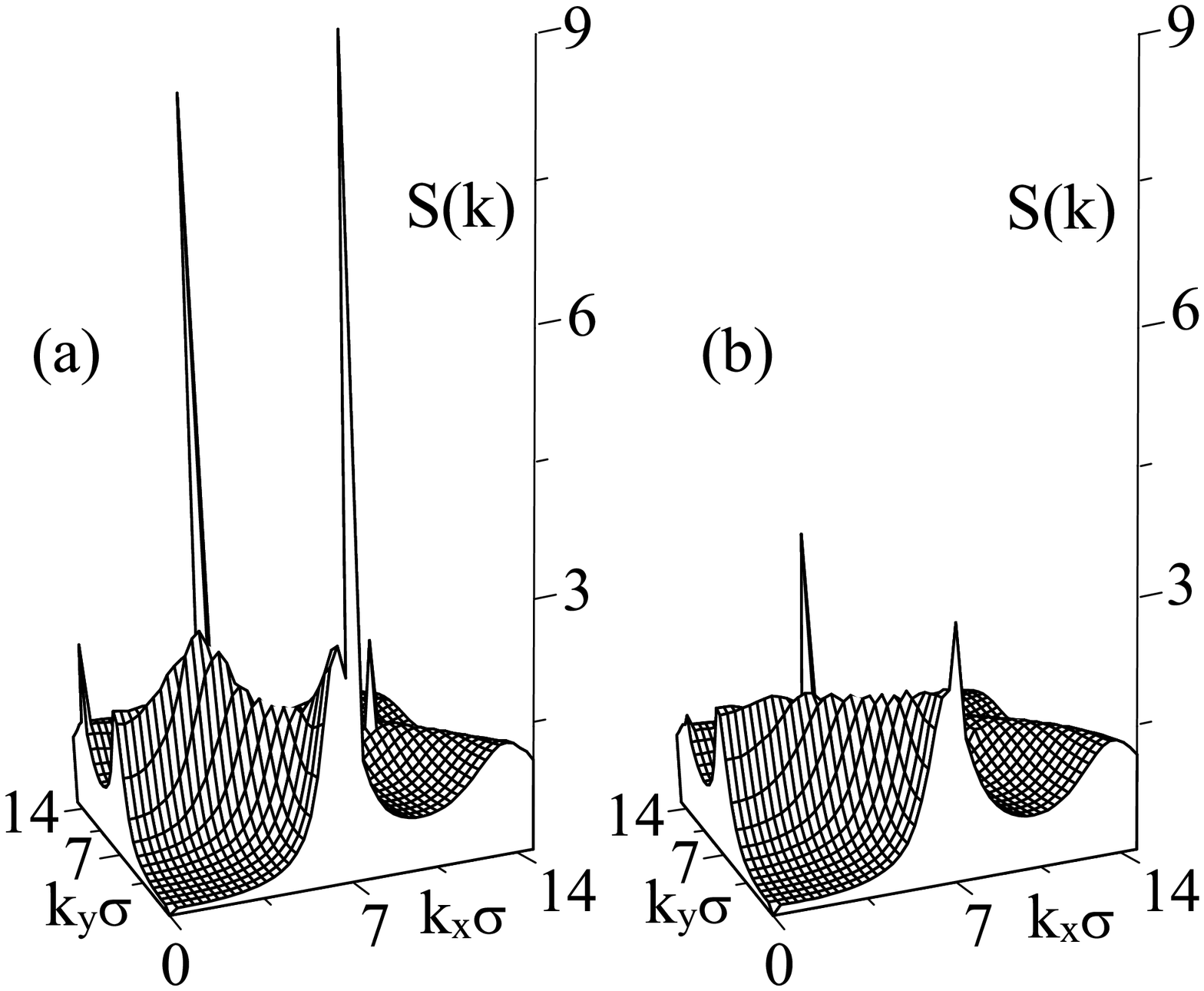}
\vspace*{-15mm}%
\caption{Structure factor $S(k_x,k_y)$ in the first fluid layer for
the fluid density $\rho\,{=}\,0.91\,\sigma^{-3}$. The upper wall
velocity is $U\,{=}\,\,0.05\,\sigma/\tau$ (a) and
$U\,{=}\,\,5.0\,\sigma/\tau$ (b). The same flow conditions as in
Fig.\,\ref{shear_velo}.} \label{SK}
\end{figure}

Figure\,\ref{SK} shows the averaged structure factor in the first
fluid layer moving with a constant velocity against the stationary
lower wall. The circular ridge at the wavevector
$|\mathbf{k}|\!\approx\!2\pi\!/\sigma$ is attributed to the
short-range ordering of the fluid monomers. The periodic potential
of the crystalline substrate induces several sharp peaks in the
fluid structure factor, which are reduced at higher slip velocities.
The first reciprocal lattice vector in the direction of shear flow
is $\mathbf{G}_1\,{=}\,(7.23\,\sigma^{-1},0)$. Although the
nearest-neighbor distance between equilibrium lattice sites in the
$xy$ plane is equal to the diameter of the LJ monomers, the surface
induced order in the first fluid layer is frustrated by the
topological constraints associated with packing of polymer chains
near the surface.

\begin{figure}[t]
\includegraphics[width=10.cm,height=7.35cm,angle=0]{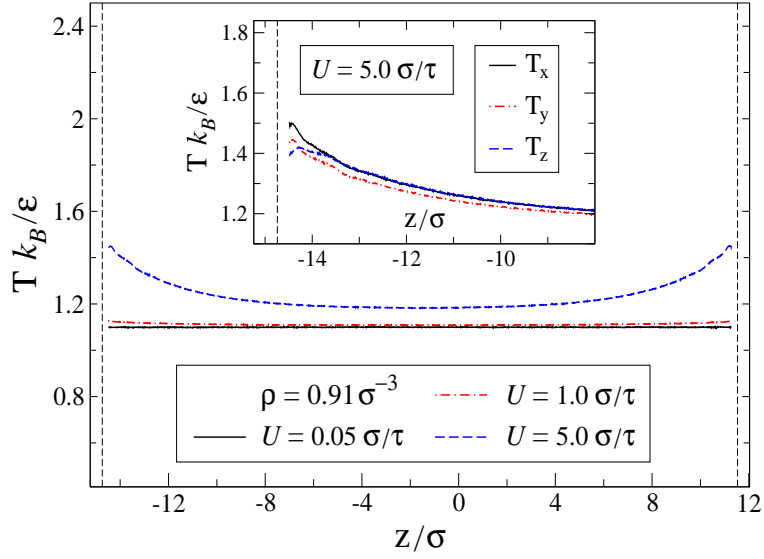}
\caption{(Color online) Temperature profiles for the indicated
values of the upper wall speed $U$ and the fluid density
$\rho\,{=}\,0.91\,\sigma^{-3}$. Vertical dashed lines denote the
position of the liquid/solid interface. The inset shows the
$\hat{x}$, $\hat{y}$ and $\hat{z}$ components of the temperature
profile near the stationary lower wall when the velocity of the
upper wall is $U\,{=}\,\,5.0\,\sigma/\tau$.} \label{temper_xyz}
\end{figure}

\begin{figure}[t]
\includegraphics[width=10.cm,height=7.35cm,angle=0]{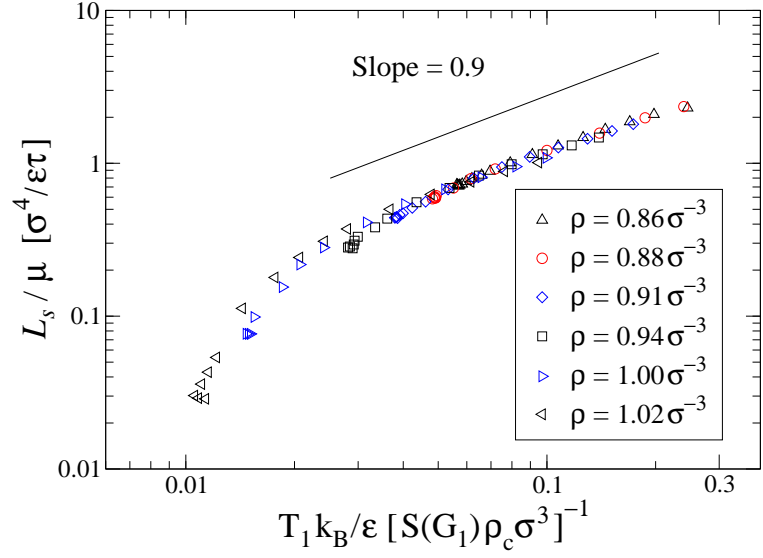}
\caption{(Color online) Log-log plot of $L_s/\mu$ as a function of
the ratio
$T_1k_B/\varepsilon\,[S(\mathbf{G}_1)\,\rho_c\,\sigma^3]^{-1}$ for
the indicated values of the polymer density. The solid line with a
slope $0.9$ is plotted for reference.}
\label{ls_div_mu_vs_T_S7_rho_c_low}
\end{figure}

In the recent MD study on shear flow of simple fluids past rigid fcc
walls~\cite{Priezjev07}, it was shown that the slip length scales
according to
\begin{equation}
L_s\!\sim[\,T_1/S(\mathbf{G}_1)\,\rho_c]^{\,1.44},%
\label{scaling}%
\end{equation}
where $\mathbf{G}_1$ is the the first reciprocal lattice vector,
$T_1$ and $\rho_c$ are the temperature and the contact density of
the first fluid layer respectively. This relation was found to hold
in a wide range of shear rates and for weak wall-fluid interactions
$0.3\!\leqslant\!\varepsilon_{\rm
wf}/\,\varepsilon\!\leqslant\!1.1$. It is expected, however, that
the scaling relation given by Eq.\,(\ref{scaling}) is not valid at
higher surface energies because the first fluid layer is locked to
the solid wall and, as a consequence, the slip length is
negative~\cite{Thompson90}. Note that the surface induced peaks in
the structure factor are much higher for polymers (see
Fig.\,\ref{SK}) than for simple fluid~\cite{Priezjev07}. Since the
viscosity is independent of shear rate for simple fluids, the
scaling relation Eq.\,(\ref{scaling}) also predicts the dependence
of the friction coefficient on microscopic parameters of the
adjacent fluid layer.

We computed the parameters entering Eq.\,(\ref{scaling}) in the
first fluid layer for the polymer melt. The fluid temperature was
estimated from the kinetic energy
\begin{equation}
k_BT=\frac{m}{3N}\sum_{i=1}^N\,[\dot{\mathbf{r}}_i-\mathbf{v}(\mathbf{r}_i)]^2,
\label{temp3d}
\end{equation}
where $\dot{\mathbf{r}}_i$ is the instantaneous velocity of the
fluid monomer and $\mathbf{v}(\mathbf{r}_i)$ is the local average
flow velocity. Averaged temperature profiles are shown in
Fig.\,\ref{temper_xyz} for the same flow conditions as in
Figs.\,\ref{mol_dens} and \ref{shear_velo}. At low shear rates
$\dot{\gamma}\lesssim0.01\,\tau^{-1}$, the fluid temperature is
equal to its equilibrium value of $T\,{=}\,1.1\,\varepsilon/k_B$
determined by the Langevin thermostat. As in the case of simple
fluids~\cite{Priezjev07,PriezjevJCP}, with further increase of the
shear rate, the polymer melt heats up and the temperature profile
becomes non-uniform across the channel. The fluid temperature is
higher near the walls because of the large slip velocity, which is
comparable to the thermal velocity, $v^2_{T}\!=k_BT\!/m$, at high
shear rates. The non-uniform temperature distribution might be
related to the slight curvature in the velocity profile for the top
wall speed $U\,{=}\,\,5.0\,\sigma/\tau$ shown in
Fig.\,\ref{shear_velo}. We can also notice that at high shear rates,
the temperature in the $\hat{y}$ direction, in which the Langevin
thermostat is applied, is slightly smaller than its value in the
$\hat{x}$ and $\hat{z}$ directions (see inset of
Fig.\,\ref{temper_xyz}). This difference implies that the kinetic
energy in the $\hat{y}$ direction dissipates faster than the energy
transfer from the other directions. In principle, it is important to
investigate whether the choice of the thermostat, e.g. dissipative
particle dynamics thermostat, has an effect on the velocity and
temperature profiles at high shear rates~\cite{Patorino07}. In the
present study, the temperature in the first fluid layer was computed
from
\begin{equation}
T_{1}=\int_{z_0}^{z_1}T(z)\rho(z)dz \,\big/
\int_{z_0}^{z_1}\rho(z)dz, \label{temper_def}
\end{equation}
where the limits of integration ($z_0=-14.4\,\sigma$ and
$z_1=-13.8\,\sigma$) are given by the width of the first peak in the
density profile (e.g. see Fig.\,\ref{mol_dens}).


\begin{figure}[t]
\includegraphics[width=10.cm,height=7.35cm,angle=0]{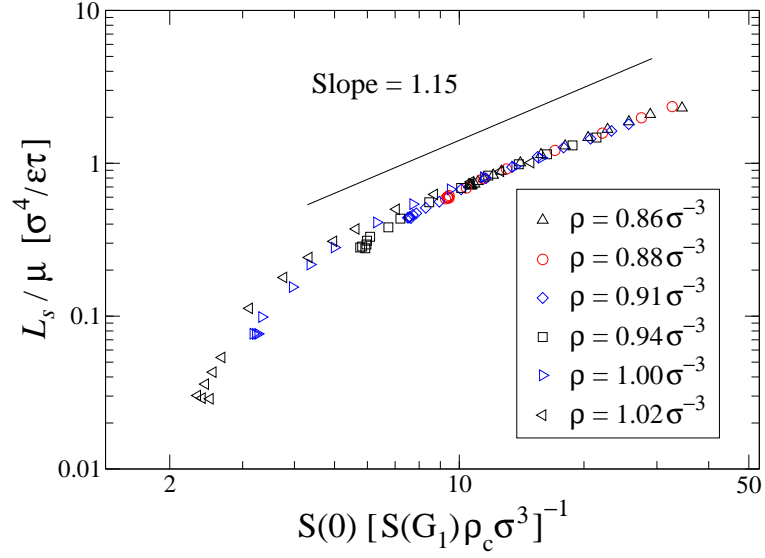}
\caption{(Color online) Log-log plot of the ratio $L_s/\mu$ as a
function of the variable
$S(0)/\,[S(\mathbf{G}_1)\,\rho_c\,\sigma^3]$. The values of the
fluid density are tabulated in the inset. The same data as in
Fig.\,\ref{ls_div_mu_vs_T_S7_rho_c_low}. The straight solid line
with a slope $1.15$ is plotted as a reference.}
\label{ls_div_mu_vs_S0_div_S7_ro_c_low}
\end{figure}

The ratio of the slip length to viscosity is plotted in
Fig.\,\ref{ls_div_mu_vs_T_S7_rho_c_low} as a function of
$T_1/[S(\mathbf{G}_1)\,\rho_c]$ for different fluid densities and
shear rates. Except for higher densities at low shear rates, the
data collapse onto a single curve, which approximately follows the
power law function with a slope of $0.9$ (see
Fig.\,\ref{ls_div_mu_vs_T_S7_rho_c_low}). Although the slope is
different from the value $1.44$ reported for simple
fluids~\cite{Priezjev07}, it is remarkable that the same combination
of parameters determines the boundary conditions for both polymer
melts and simple fluids. The simulation results also show that for
each value of the fluid density, the number of fluid monomers in the
first layer $N_{\ell}\,{=}\,S(0)$ decreases by about $4\%$ at the
highest shear rates. The same data for the slip length and viscosity
are replotted as a function of the variable
$S(0)/\,[S(\mathbf{G}_1)\,\rho_c]$ in
Fig.\,\ref{ls_div_mu_vs_S0_div_S7_ro_c_low}. The collapse of the
data is consistent with the previous results for the
rate-independent slip length at the interface between simple fluids
and crystalline walls~\cite{Thompson90}. The correlation between the
friction coefficient and the fluid structure in the adjacent layer
presented in Figs.\,\ref{ls_div_mu_vs_T_S7_rho_c_low}
and~\ref{ls_div_mu_vs_S0_div_S7_ro_c_low} suggests a possible
existence of a general functional relation between the slip length,
shear rate and polymer density.

\section{Conclusions}

In this paper, the rate dependence of the slip length in the shear
flow of polymer melts past atomically smooth, thermal surfaces was
studied using molecular dynamics simulations. For weak wall-fluid
interactions, the slip length passes through a minimum as a function
of shear rate and then increases rapidly at higher shear rates. The
nonlinear rate dependence of the slip length was analyzed in terms
of the dynamic friction at the liquid/solid interface. In a wide
range of fluid densities, the friction coefficient (the ratio of the
shear viscosity and the slip length) undergoes a universal
transition from a constant value to the power law decay as a
function of the slip velocity. The simulation results for polymer
melts confirm the previous findings for simple
fluids~\cite{Priezjev07} that the surface induced structure and the
contact density of the adjacent fluid layer are crucial factors for
determining the value of the friction coefficient.

Future research will be focused on the rate dependence of the slip
length for higher fluid densities and entangled polymer melts.

\section*{Acknowledgments}
Financial support from the Michigan State University Intramural
Research Grants Program is gratefully acknowledged. Computational
work in support of this research was performed at Michigan State
University's High Performance Computing Facility.

\bibliographystyle{prsty}

\begin{thebibliography}{80}


\bibitem{MicroNano}   L. Bocquet and J.-L. Barrat, Soft Matter {\bf 3}, 685 (2007).

\bibitem{Churaev84}   N.~V. Churaev, V.~D. Sobolev, and A.~N. Somov, J. Colloid Interface Sci. {\bf 97}, 574 (1984). 

\bibitem{LegerPRL93}  K.~B. Migler, H. Hervet, and L. Leger, Phys. Rev. Lett. {\bf 70}, 287 (1993).

\bibitem{Charlaix01}  J. Baudry, E. Charlaix, A. Tonck, and D. Mazuyer, Langmuir {\bf 17}, 5232 (2001).

\bibitem{Archer98}    V. Mhetar and L.~A. Archer, Macromolecules {\bf 31}, 6639 (1998).

\bibitem{MackayVino}  R.~G. Horn, O.~I. Vinogradova, M.~E. Mackay, and N. Phan-Thien,
                      J. Chem. Phys. {\bf 112}, 6424 (2000). 

\bibitem{Mackay00}    K.~M. Awati, Y. Park, E. Weisser, and M.~E. Mackay,
                      J. Non-Newton. Fluid Mech. {\bf 89}, 117 (2000).

\bibitem{Granick01}   Y. Zhu and S. Granick, Phys. Rev. Lett. {\bf 87}, 096105 (2001). 

\bibitem{Breuer03}    C.~H. Choi, K.~J.~A. Westin, and K.~S. Breuer, Phys. Fluids {\bf 15}, 2897 (2003). 

\bibitem{Archer03}    J. Sanchez-Reyes and L.~A. Archer, Langmuir {\bf 19}, 3304 (2003). 

\bibitem{Granick02}   Y. Zhu and S. Granick, Phys. Rev. Lett. {\bf 88}, 106102 (2002).

\bibitem{Vinograd06}  O.~I. Vinogradova and G.~E. Yakubov, Phys. Rev. E {\bf 73}, 045302(R) (2006). 

\bibitem{Leger06}     T. Schmatko, H. Hervet, and L. Leger, Langmuir {\bf 22}, 6843 (2006).

\bibitem{KB89}        J. Koplik, J.~R. Banavar, and J.~F. Willemsen, Phys. Fluids A {\bf 1}, 781 (1989).

\bibitem{Thompson90}  P.~A. Thompson and M.~O. Robbins, Phys. Rev. A {\bf 41}, 6830 (1990). 

\bibitem{Barrat99fd}  J.-L. Barrat and L. Bocquet, Faraday Discuss. {\bf 112}, 119 (1999). 

\bibitem{Barrat99}    J.-L. Barrat and L. Bocquet, Phys. Rev. Lett. {\bf 82}, 4671 (1999). 

\bibitem{Tanner99}    A. Jabbarzadeh, J.~D. Atkinson, and R.~I. Tanner, J. Chem. Phys. {\bf 110}, 2612 (1999). 

\bibitem{Cieplak01}   M. Cieplak, J. Koplik, J.~R. Banavar, Phys. Rev. Lett. {\bf 86}, 803 (2001).

\bibitem{Priezjev04}  N.~V. Priezjev and S.~M. Troian, Phys. Rev. Lett. {\bf 92}, 018302 (2004). 

\bibitem{Attard04}    T.~M. Galea and P. Attard, Langmuir {\bf 20}, 3477 (2004). 

\bibitem{Priezjev05}  N.~V. Priezjev, A.~A. Darhuber, and S.~M. Troian, Phys. Rev. E {\bf 71}, 041608 (2005). 

\bibitem{Priezjev07}  N.~V. Priezjev, Phys. Rev. E {\bf 75}, 051605 (2007). 

\bibitem{Thompson95}  P.~A. Thompson, M.~O. Robbins, and G.~S. Grest,
                      Israel Journal of Chemistry {\bf 35}, 93 (1995). 

\bibitem{dePablo96}   R. Khare, J.~J. de\,Pablo, and A. Yethiraj, Macromolecules {\bf 29}, 7910 (1996). 

\bibitem{Koike98}     A. Koike and M. Yoneya, J. Phys. Chem. B {\bf 102}, 3669 (1998).

\bibitem{Andrienko05} X. Zhou, D. Andrienko, L. Delle Site, and K. Kremer, J. Chem. Phys. {\bf 123}, 104904 (2005).

\bibitem{Binder06}    C. Pastorino, K. Binder, T. Kreer, and M. Muller, J. Chem. Phys. {\bf 124}, 064902 (2006).

\bibitem{Barsky01}    S. Barsky and M.~O. Robbins, Phys. Rev. E {\bf 63}, 021801 (2001).

\bibitem{Nature97}    P.~A. Thompson and S.~M. Troian, Nature (London) {\bf 389}, 360 (1997). 

\bibitem{Bird87}      R.~B. Bird, C.~F. Curtiss, R.~C. Armstrong, and O. Hassager,
                      {\it Dynamics of Polymeric Liquids} 2nd ed. (Wiley, New York, 1987). 

\bibitem{Kremer90}    K. Kremer and G.~S. Grest, J. Chem. Phys. {\bf 92}, 5057 (1990). 

\bibitem{Grest86}     G.~S. Grest and K. Kremer, Phys. Rev. A {\bf 33}, 3628 (1986). 

\bibitem{GrestJCP04}  M. Tsige and G.~S. Grest, J. Chem. Phys. {\bf 120}, 2989 (2004). 

\bibitem{Allen87}     M.~P. Allen and D.~J. Tildesley,
                      {\it Computer Simulation of Liquids} (Clarendon, Oxford, 1987). 

\bibitem{Barrat03}    J.~L. Barrat and J.~P. Hansen, {\it Basic concepts for simple and complex liquids}
                      (Cambridge University Press, Cambridge, 2003). 

\bibitem{PriezjevJCP} N.~V. Priezjev, J. Chem. Phys. {\bf 127}, 144708 (2007).

\bibitem{Israel92}    J.~N. Israelachvili, {\it Intermolecular and Surface Forces}, 
                      2nd ed. (Academic Press, San Diego, 1992).

\bibitem{Robbins97}   M.~O. Robbins, Nature (London) {\bf 389}, 331 (1997).

\bibitem{dePablo95}   Z. Xu, J.~J. de\,Pablo, and S. Kim, J. Chem. Phys. {\bf 102}, 5836 (1995). 

\bibitem{Todd04}      J.~T. Bosko, B.~D. Todd, and R.~J. Sadus, J. Chem. Phys. {\bf 121}, 12050 (2004).

\bibitem{XudePablo97} Z. Xu, R. Khare, J.~J. de\,Pablo, and S. Kim, J. Chem. Phys. {\bf 106}, 8285 (1997).

\bibitem{Priezjev08}  A. Niavarani and N.~V. Priezjev, unpublished (cond-mat/0801.3828).

\bibitem{Smith96}     E.~D. Smith, M.~O. Robbins, and M. Cieplak, Phys. Rev. B {\bf 54}, 8252 (1996).

\bibitem{Tomassone97} M.~S. Tomassone, J.~B. Sokoloff, A. Widom, J. Krim, Phys. Rev. Lett. {\bf 79}, 4798 (1997).

\bibitem{Patorino07}  C. Pastorino, T. Kreer, M. Muller, and K. Binder, Phys. Rev. E {\bf 76}, 026706 (2007).












\end{thebibliography}

\end{document}